\documentclass[prl,twocolumn,showpacs,floatfix,superscriptaddress]{revtex4-1}

\usepackage{amssymb,amsmath,graphicx}
\usepackage{color}
\begin{document}

\title{\boldmath Landau Level Spectroscopy of Dirac Electrons in a Polar Semiconductor with Giant Rashba Spin Splitting \unboldmath}

\author{S\'andor Bord\'acs}
\affiliation{Quantum Phase Electronics Center and Department of Applied Physics, University of Tokyo, Tokyo 113-8656, Japan}

\author{Joseph G. Checkelsky}
\affiliation{Quantum Phase Electronics Center and Department of Applied Physics, University of Tokyo, Tokyo 113-8656, Japan}
\affiliation{RIKEN Center for Emergent Matter Science (CEMS), Wako 351-0198, Japan}

\author{Hiroshi Murakawa}
\affiliation{RIKEN Center for Emergent Matter Science (CEMS), Wako 351-0198, Japan}

\author{Harold Y. Hwang}
\affiliation{RIKEN Center for Emergent Matter Science (CEMS), Wako 351-0198, Japan}
\affiliation{Stanford Institute for Materials and Energy Sciences, SLAC National Accelerator Laboratory, Menlo Park, California 94025, USA}

\author{Yoshinori Tokura}
\affiliation{Quantum Phase Electronics Center and Department of Applied Physics, University of Tokyo, Tokyo 113-8656, Japan}
\affiliation{RIKEN Center for Emergent Matter Science (CEMS), Wako 351-0198, Japan}

\date{\today}

\begin{abstract}
Optical excitations of BiTeI with large Rashba spin splitting have been studied in an external magnetic field ($B$) applied parallel to the polar axis. A sequence of transitions between the Landau levels (LLs), whose energies are in proportion to $\sqrt{B}$ were observed, being characteristic of massless Dirac electrons. The large separation energy between the LLs makes it possible to detect the strongest cyclotron resonance even at room temperature in moderate fields. Unlike in 2D Dirac systems, the magnetic field induced rearrangement of the conductivity spectrum is directly observed.
\end{abstract}

\pacs{
}

\maketitle
Helical Dirac electrons have recently attracted much attention in the field of spintronics owing to their interlocked spin and translational momentum \cite{Brune2012,Pesin2012,Garate2010}. The materials with these massless electrons can work as a spin-polarized current generator, and hence may find applications also in the field of spin-optoelectronics \cite{McIver2011,Raghu2010,Hosur2011}. The criteria for the emergence of such a helical Dirac cone are the lack of inversion symmetry and the presence of spin-orbit coupling; they are fulfilled most typically for the surface electronic states of the topological insulators as confirmed by recent experiments \cite{Hsieh2008,Xia2009}.

One of the most intriguing phenomena related to massless electrons is their unusual dynamics in an external magnetic field which was first observed as a half-integer shift of the quantum Hall plateaus of graphene \cite{Novoselov2005,Zhang2005}. This result was interpreted in terms of the unique Landau level (LL) spectrum of Dirac fermions: $E_{\pm N}$=$\pm v_F\sqrt{2e\hbar BN}$, which has a magnetic field independent level at $E_0$=0 in the presence of electron-hole symmetry. Here $v_F$ is the Fermi velocity, $e$ is the electron charge, and $\hbar$ is the Planck constant divided by 2$\pi$. The energy levels have a square root dependence on both the external magnetic field $B$ and LL index $N$, which is in stark contrast to the usual linear $B$ and $N$ dependence of parabolic bands, as it was confirmed by tunneling \cite{Miller2009} and infrared spectroscopy \cite{Sadowski2006,Jiang2007,Orlita2008} in graphene. Moreover, the Fermi velocity $v_F$ was determined accurately by these techniques. Another fingerprint of the Dirac electrons is that the electromagnetic radiation can induce several transitions between the unequally spaced LLs governed by the selection rules: $\pm N\rightarrow$$\pm$$|N\pm$1$|$ \cite{Sadowski2006,Jiang2007}. However, the LL quantization of electronic states on a non-degenerate, helical Dirac cone has thus far only been demonstrated on the surface of topological insulators \cite{Cheng2010,Hanaguri2010,Schafgans2012}. In this Letter, we report the LL spectroscopy of massless Dirac electrons arising from Rashba spin split bands. The giant Rashba-type spin splitting in BiTeI \cite{Ishizaka2011} enables us to detect a sequence of LL transitions of the quantized massless Dirac cone in a bulk material for the first time.


\begin{figure}[h!]
\includegraphics[width=3.3in]{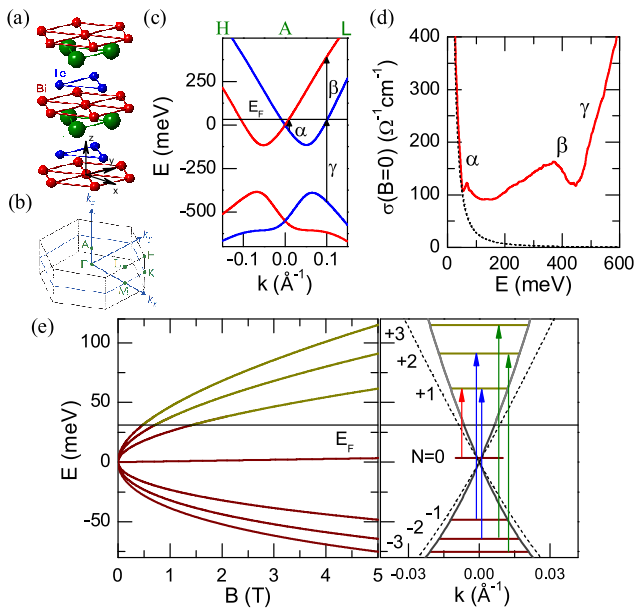}
\caption{(color online). Band structure and Landau-level formation in BiTeI. (a) The non-centrosymmetric crystal structure, and (b) the Brillouin zone of BiTeI. (c) The bulk in-plane electron dispersion relation as determined by relativistic, first-principles calculations \cite{Ishizaka2011,Bahramy2011}. The red and blue curves correspond to the "spin-up" and the "spin-down" bands, respectively. (d) Optical conductivity spectrum of BiTeI measured at T=10\,K. Intra- and inter-band transitions are labeled according to the convention used in panel (c). (e) The magnetic field dependence of the lowest LLs in a Rashba model, which describes the levels in vicinity of the Dirac point of the conduction band, are depicted on the left-hand side. The thin solid line indicates the position of the Fermi energy in the present crystals. The inter-LL transitions in $B$=5\,T allowed by optical selection rules are plotted on the right-hand side together with the zero-field electron dispersion determined by the first-principles calculations (thin dashed curves) and within the Rashba model (thick solid curves).}
\label{Fig1}
\end{figure}

The non-centrosymmetric crystal structure of BiTeI ({\it P3m1}), shown in Fig.~\ref{Fig1}(a), consists of alternating stacks of Bi, Te and I layers. According to recent angular-resolved photoemission spectroscopy (ARPES) \cite{Ishizaka2011,Crepaldi2012,Landolt2012,Sakano2012}, optical spectroscopy \cite{Lee2011,Demko2012}, magneto-transport \cite{Martin2013,Bell2013} studies and relativistic first-principles calculations \cite{Bahramy2011,Eremeev2012}, the polar structure and the large spin-orbit interaction give rise to giant Rashba-type spin splitting in the bulk conduction and valance bands of this material (Fig.~\ref{Fig1}(c)). In the vicinity of the Dirac point, which is the intersection of the spin-polarized conduction bands protected by time-reversal symmetry, the in-plane energy-momentum dispersion and the spin texture are approximated by a simple Rashba model:
\begin{equation}
H=\frac{p^2}{2m^*}+\frac{\alpha}{\hbar}{\boldsymbol\sigma}\cdot({\bf\hat{z}}\times{\bf p}),
\label{EqRashba}
\end{equation}
where $m^*$$\approx$$0.09m_o$ is the effective mass in the unit of the free electron mass $m_o$, $\alpha$ is the Rashba parameter, {\bf\^{z}} is the unit vector parallel to the [001] direction, {\bf p} and {$\boldsymbol\sigma$} are the momentum and the spin operators, respectively. The second term, responsible for the momentum $k$-linear dispersion as well as for the spin splitting, leads to the appearance of the helical Dirac cone in BiTeI as in topological insulators \cite{Liu2010}. The bulk band structure parameters, the Rashba parameter $\alpha$ and the effective mass $m^*$ of the Rashba model, are taken from the relativistic first-principles calculations of M.S. Bahramy et al.\cite{Bahramy2011} which are consistent with a broad variety of experiments: ARPES \cite{Ishizaka2011,Sakano2012,Sakano2013}, optical spectroscopy \cite{Lee2011,Demko2012} and Shubnikov--de Haas oscillations \cite{Bell2013}. The velocity of the electrons on the inner Fermi surface near the Dirac point is approximately $v$$_F$=$\frac{\alpha}{\hbar}$$\approx$6.61$\times$10$^5$\,m/s in BiTeI as estimated within the Rashba model. In contrast to the case of graphene where the Fermi velocity is fixed by the overlap of the carbon 2$p$ orbitals, in a Rashba system the opening angle of the cone can be modified more easily either by tuning the spin-orbit interaction or by varying the asymmetry in the potential, as observed in a series of sister compounds BiTe$X$ ($X$=I, Br, Cl) \cite{Sakano2013}.

When an external magnetic field is applied along the polar axis, the in-plane motion of the electrons becomes quantized to form LLs. Following E.~I.~Rashba \cite{Rashba1960}, the spectrum of the LLs can be calculated using the band structure parameters of BiTeI as shown in Fig.~\ref{Fig1}(e). The electron dispersion along the z direction (not indicated in the figure) alters the $\delta$-function like anomalies in the density of states to the $\sim$1/$\sqrt{E}$ van-Hove singularities without changing the position of the resonances \cite{Rashba1960}. Since in BiTeI the Rashba energy $E_R$=$\frac{1}{2}m^*v_F^2$$\sim$113\,meV is much larger than the cyclotron frequency $\omega_c$=$\frac{eB}{m^*}$ ($\sim$1.3\,meV for $B$=1\,T) in the present experimental field range, the energy spectrum for the LLs with low index ($N$) is dominated by the contribution of the massless Dirac electrons: $E$$_{\pm N}\approx\pm v_F\sqrt{2e\hbar BN}$+$\hbar\omega_c$$N$ and $E$$_0$=$\frac{\hbar\omega_c}{2}$, where the Zeeman effect is neglected. Furthermore, the spin-orbit coupling mixes the spin states and the cyclotron orbits, which leads to a completely spin-polarized 0$^{th}$ LL and electric-dipole allowed inter-LL transitions accompanying spin flip \cite{Rashba1960,LLspectroscopy1991}. Such a combined resonance has been observed in semiconductors with large spin-orbit coupling, e.g.~InSb \cite{Palik1970}. The selection rule for the optical excitations between the LLs in the Rashba model is identical to that known for Dirac systems: $\pm$$N$$\rightarrow$$\pm$$|$$N$$\pm$1$|$. Therefore, when the Fermi energy is close to the Dirac point, the inner Fermi surface electrons of a Rashba system with large spin-orbit coupling follow the same low-energy dynamics as helical Dirac electrons up to moderate magnetic fields. The major optical transitions at $B$=5\,T expected in the vicinity of the Dirac point are plotted in Fig.~\ref{Fig1}(e).

Optical experiments were performed on the cleaved $ab$-plane of BiTeI single crystals which were grown by the Bridgman method (details are given elsewhere \cite{Ishizaka2011}). Near-normal incidence reflectivity was measured at various temperatures in the energy region $E$=0.024-0.85\,eV. To perform an accurate Kramers-Kronig analysis, the reflectivity spectra at zero field and room temperature were measured up to 40\,eV with use of synchrotron radiation at UV-SOR, Institute for Molecular Science, Japan \cite{Supplement}. The intensity ratio of the infrared spectra at finite and zero magnetic field was measured with a Fourier transform infrared (FT-IR) spectrometer coupled to a room-temperature bore magnet (Oxford Microstat BT) in the Faraday geometry where both the magnetic field and the light propagation direction were parallel to the $c$-axis. The small magnetic field dependence of the baseline in the reflectivity ratios were taken into account by normalizing the spectra in the featureless energy region above 170\,meV \cite{Supplement}. Optical conductivity spectra were obtained from the absolute reflectivity spectra by Kramers-Kronig transformation.

The position of the Fermi energy for the sample used in the present study was determined and the band structure parameters were verified from the optical conductivity spectra measured at T=10\,K (Fig.~\ref{Fig1}(d)). In agreement with the former optical studies \cite{Lee2011,Demko2012}, below the band gap ($\gamma$-transition) we observed a broad continuum of the intra-band excitations in addition to the Drude peak of free carriers dominating the lowest energy range. This broad hump accompanied by the van Hove singularity features ($\alpha$ and $\beta$, see Fig.~\ref{Fig1}(c)) is characteristic of the intra-band excitations within the spin-split Rashba bands allowed by the strong spin-orbit coupling. The Fermi energy, estimated from the energy of the $\alpha$-transition ($E_\alpha$), is $E_F$=$E_\alpha$/2$\approx$30\,meV above the Dirac point. This value is in accordance with the carrier concentration, $n$$\sim$4$\times$10$^{19}$cm$^{-3}$, deduced from the measured Hall coefficient.


\begin{figure}[h!]
\includegraphics[width=2.8in]{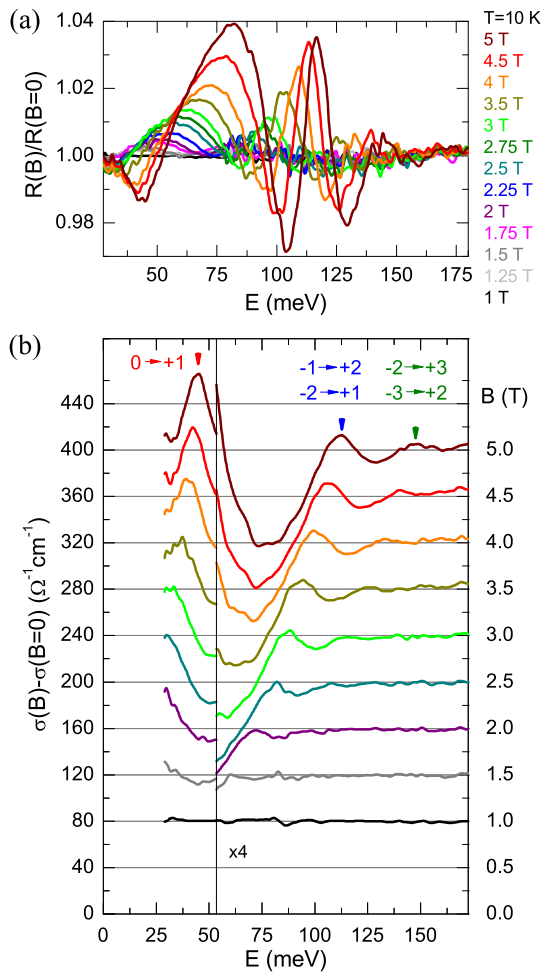}
\caption{(color online). Relative reflectivity and optical conductivity spectra at $T$=10\,K. (a) Magnetic field induced changes in the mid-infrared reflectivity spectra. (b) Magnetic field dependence of the relative optical conductivity spectra. The spectra are shifted vertically in proportion to the magnetic field $B$. The three peak structures assigned to the inter-LL transitions are marked with arrows.}
\label{Fig2}
\end{figure}

Magnetic-field dependent oscillations were observed in the reflectivity spectra at $T$=10\,K as plotted relative to the zero field spectrum in Fig.~\ref{Fig2}(a). To determine the energy of the transitions between the LLs, the field induced changes in the optical conductivity spectra were calculated by the Kramers-Kronig transformation. First, we analyze the relative optical conductivity spectrum measured at $B$=5\,T (see the topmost curve of Fig.~\ref{Fig2}(b)). Three peaks can be clearly discerned in the spectrum and assigned to the lower-energy inter-LL transitions as shown in Fig.~\ref{Fig1}(e). These excitations cannot be attributed to the weak dielectric response of surface states since the corresponding changes in the relative reflectivity spectra, mostly appearing below the plasma edge $E_{pl}\approx$110\,meV, are as large as 4$\%$ in $B$=5\,T \cite{Supplement}. The magnitude of the magneto-reflectance oscillations is rather comparable to that of detected for the bulk inter-LL transitions in elemental bismuth \cite{Vecchi1976}. Furthermore, the inter-LL transitions involving the low-index LLs can be detected only when the Fermi-energy is close to the Dirac point, which is the case for the bulk states of the present sample. Since the Dirac point of the surface states is shifted by $\sim\pm$250\,meV from the bulk Fermi-energy at either the I or the Te terminated surfaces \cite{Crepaldi2012,Sakano2013}, the observed excitations can only be assigned to the bulk electronic states.

Theoretically, the parabolic feature of the band structure is expected to cause a splitting proportional to the cyclotron energy, $\pm\hbar\omega_c$ for the transitions -$N$$\rightarrow$$N$+1 and -$N$-1$\rightarrow$$N$, and $\pm\frac{1}{2}\hbar\omega_c$ for the excitations 0$\rightarrow$+1 and -1$\rightarrow$0. The transition with the lowest energy, 0$\rightarrow$+1 at $B$=5\,T, is always non-degenerate because of Pauli blocking. However, no parabolicity-induced splitting was observed in the present experiment even for the higher energy peaks. This may indicate that the $k$-linear dispersion around the Dirac point is more robust in BiTeI compared to the prediction of a simple Rashba model, which is also supported by the results of first-principles calculation (see and compare thin dashed and thick solid lines in the right panel of Fig.~\ref{Fig1}(e)). The detailed magnetic-field dependence of the optical conductivity spectra is shown in Fig.~\ref{Fig2}(b). The inter-LL transitions were observed above $B$$>$1.5\,T; with increasing magnetic field all the resonances shift toward higher energies and their oscillator strengths increase as expected.

\begin{figure}[h!]
\includegraphics[width=3.2in]{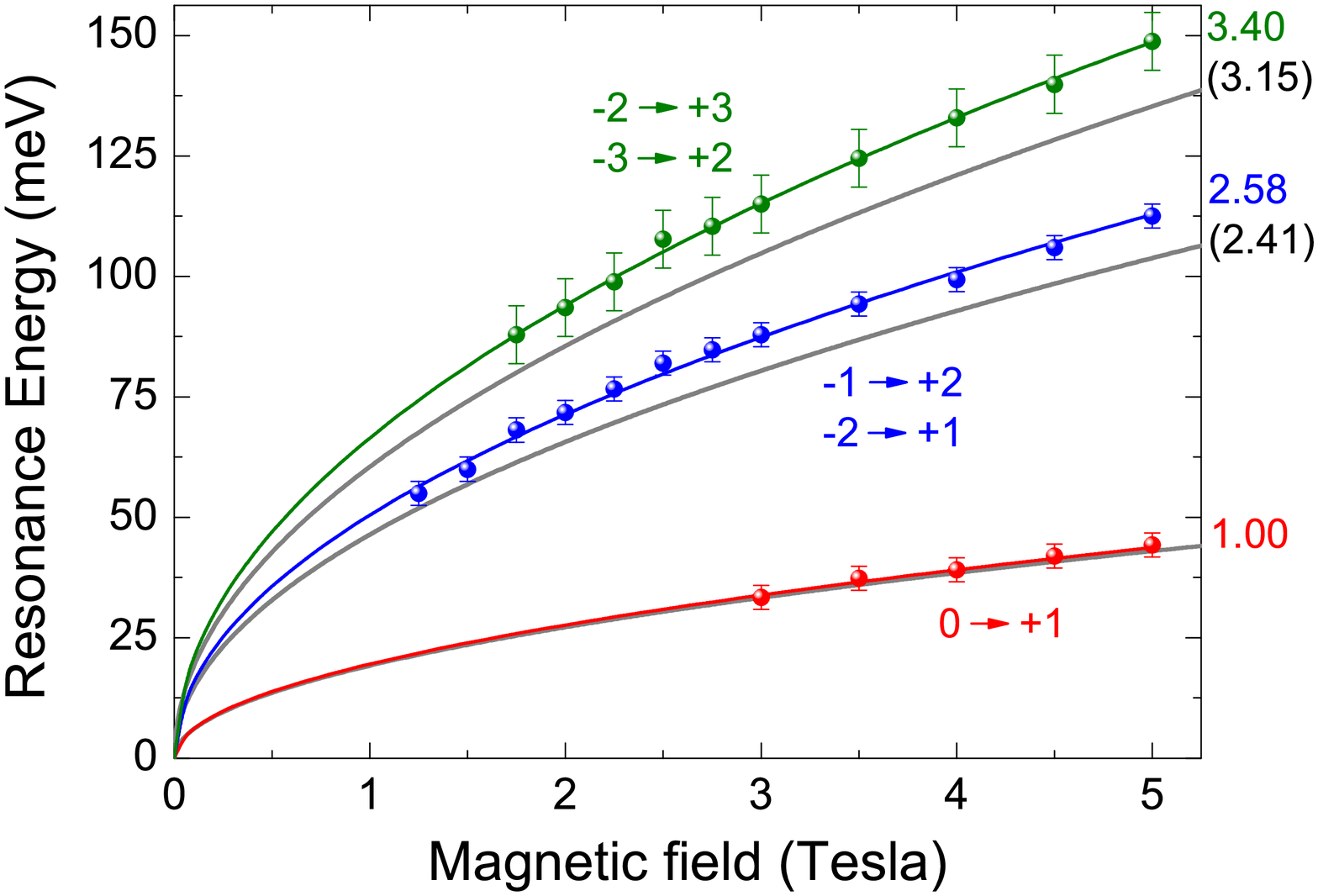}
\caption{(color online). Magnetic field dependence of the inter-LL transitions. The resonance energies of the three observed inter-LL excitation branches (colored balls) can be fitted with $\sqrt{B}$ field dependence (colored solid curves). The excitation energies determined from the experiment approximately follow those calculated from the Dirac model (gray curves). The fitted curves scales as 1:2.58:3.40 which is close to the ratios of the inter-LL transition energies for Dirac electrons 1:1+$\sqrt{2}$:$\sqrt{2}$+$\sqrt{3}$.}
\label{Fig3}
\end{figure}

Although the line shape of the relative conductivity spectra cannot be simply decomposed to a sum of Lorentzians, the positions of the three maxima are identified as the resonance energies, which are plotted as a function of the external magnetic field in Fig.~\ref{Fig3}. All the three branches can be well fitted with $\sqrt{B}$ field dependence (see colored lines in Fig.~\ref{Fig3}). The Fermi velocity $v$$_F^{(exp)}$=5.35$\times$10$^5$\,m/s, deduced from the 0$\rightarrow$+1 transition, is about $\sim$80$\%$ of the velocity calculated from the Rashba parameter. However, the Fermi velocity derived directly from the first principles calculations, $v$$_F^{(calc)}$=5.3$\times$10$^5$\,m/s is in good agreement with the value observed in the present experiment, implying that the electron dispersion can be described by the massless Dirac electron model in even a wider energy and magnetic field range. The fitted curves scale as 1:2.58:3.40, respectively, which is close to that expected for the ratio of the inter-LL transitions in a Dirac system; 1:1+$\sqrt{2}$($\approx$2.41):$\sqrt{2}$+$\sqrt{3}$($\approx$3.15). The agreement is even better between the experiment and the theory when the resonance energies of the second and the third modes are compared; 1.32 (experiment) and $\frac{\sqrt{2}+\sqrt{3}}{1+\sqrt{2}}\approx$1.30 (theory), respectively. Nonlinear magnetic field dependence of the LL resonances, as affected by non-parabolicity due to the strong spin-orbit interaction, were observed e.g.~in the semiconductor InSb \cite{Palik1970}, and in the semimetal Bi and Bi$_{1-x}$Sb$_x$ \cite{Palik1970,Vecchi1976} with massive Dirac electrons \cite{Hsieh2008}. However, such a sequence of LL transitions characteristic of massless Dirac electrons as observed here has not been reported in bulk materials so far. As compared with the inter-LL excitation energy calculated by the Dirac model assuming v$_F^{(calc)}$=5.3$\times$10$^5$\,m/s (light gray lines in Fig.~\ref{Fig3}), a small deviation to higher energy is discerned for the second and the third modes. The negative components (around $\sim$75\,meV and $\sim$130\,meV) in the relative optical conductivity spectra may explain such a slight discrepancy, since they can cause the apparent shift of the maximum position from the transition energy.


\begin{figure}[h!]
\includegraphics[width=3in]{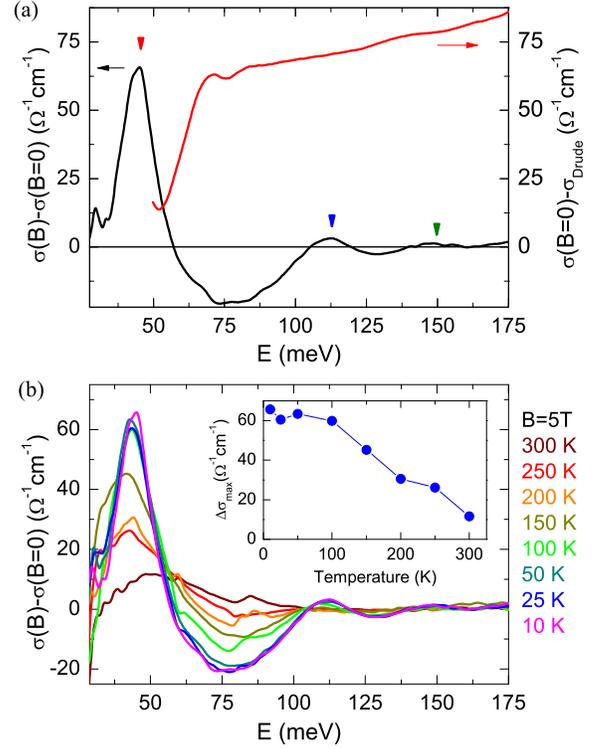}
\caption{(color online). (a) The relative optical conductivity in $B$=5\,T (black curve) and the zero-field optical conductivity spectrum after the subtraction of the Drude part (red curve). The spectral weight transfer to the inter-LL excitations leads to strong suppression of the intra-band spin-split transitions around $\sim$75\,meV. (b) Temperature dependence of the relative optical conductivity spectra in $B$=5\,T. The cyclotron resonance 0$\rightarrow$+1 was observed even at room temperature. The inset shows the temperature dependence of the maximal value $\Delta\sigma_{max}$ (around 40\,meV) of the magneto-conductivity.}
\label{Fig4}
\end{figure}

According to the sum rule for the optical conductivity, the integral of magnetic field induced changes over the entire spectrum should be zero. Thus, the accumulation of the spectral weight around the inter-LL transitions should be accompanied by negative regions in the relative optical conductivity, which was readily observed in Fig.~\ref{Fig2}(b) and Fig.~\ref{Fig4}(a). The reduction of the spectral weight is most appreciable, exceeding 40$\%$ of the intraband transitions, around $\sim$75\,meV at $B$=5\,T, where the $\alpha$- and $\beta$-transitions between the Rashba spin-split bands dominate the absorption in zero magnetic field. Such a significant spectral weight transfer from the excitations of bulk electronic states \cite{Lee2011,Demko2012} further supports the presence of bulk Dirac electrons. The Drude weight, which arises from both the inner and outer Fermi surfaces, is also expected to be transferred to the inter-LL transitions, but it was not observed here because of experimental limitations of the energy window. Contrary to the case of a usual parabolic band where only the Drude weight is transferred to the cyclotron resonance, the spectral weight of excitations between the LLs in BiTeI is likely to originate from the intraband transitions within the Rashba spin-split bands.


The temperature dependence of the magneto-conductivity spectra is shown in Fig.~\ref{Fig4}(b). No considerable change in the conductivity spectra was discerned as the temperature increased from 10\,K up to 100\,K owing to the large energy separation between the LLs. Towards elevated temperatures, the magnitude of the transitions smoothly decreases and the peaks become broader; however we can observe the 0$\rightarrow$+1 transition even at room temperature. These temperature-robust features are also indicative of the Dirac LL structure, as previously observed in graphene \cite{Orlita2008}.

In conclusion we have studied the LL spectrum of the Rashba semiconductor BiTeI by infrared spectroscopy. A sequence of inter-LL transitions characteristic of massless Dirac electrons are observed when the Fermi energy is located in the vicinity of the Dirac point. The helical spin-polarization of such a non-degenerate Dirac cone in a 3D material makes BiTeI a promising candidate to host enhanced magneto-optical effects, such as optical spin-current induction \cite{Fuseya2012}, circular photogalvanic and spin-galvanic effects \cite{Hosur2011,Ganichev2002} due to the strong optical absorption and bulk current flow.

\begin{acknowledgements}
We are grateful to N.~Nagaosa, M.S.~Bahramy, and I. K\'ezsm\'arki for fruitful discussions. This work was supported by the Funding Program for World-Leading Innovative R\&D on Science and Technology (FIRST Program), Japan. Y.T. acknowledges support by JSPS Grant-in-Aid for Scientific Research(S) No. 24224009. H.Y.H. acknowledges support from the U.S. department of Energy, the Office of Basic Energy Sciences, Materials Sciences and Engineering Division, under Contract No. DE-AC02-76SF00515.
\end{acknowledgements}

\end{document}